\documentclass[aps,prb,preprint,superscriptaddress]{revtex4-2}
\usepackage{graphicx}
\usepackage{amssymb}
\usepackage{amsmath}
\usepackage{multirow}
\usepackage{xcolor}
\usepackage{array}
\usepackage{textcomp}

\begin{document}

\title{Prediction of new superconducting bilayers heterostructures using quantum confinement and proximity effects}

\author{Giovanni Alberto Ummarino}
\affiliation{Istituto di Ingegneria e Fisica dei Materiali,
Dipartimento di Scienza Applicata e Tecnologia, Politecnico di
Torino, Corso Duca degli Abruzzi 24, 10129 Torino, Italy. ORCID 0000-0002-6226-8518}
\affiliation{National Research Nuclear University MEPhI (Moscow Engineering
Physics Institute), Kashirskoe shosse 31, Moscow 15409, Russia}
\email{giovanni.ummarino@polito.it}

\author{Alessio Zaccone}
\affiliation{Department of Physics ``A. Pontremoli'', University of Milan, via Celoria 16,
20133 Milan, Italy}
\affiliation{Institut f{\"u}r Theoretische Physik, University of G{\"o}ttingen,
Friedrich-Hund-Platz 1,
37077 G{\"o}ttingen, Germany}
\email{alessio.zaccone@unimi.it}

\begin{abstract}
A central challenge in nanoscale superconductivity is to understand and exploit the combined action of quantum confinement and proximity effects in experimentally realistic metallic heterostructures. We theoretically investigate superconducting bilayer heterostructures in which these two effects coexist. Using a generalized Eliashberg framework that incorporates both quantum confinement and proximity coupling, we show that their interplay can substantially enhance the superconducting critical temperature. In particular, the theory predicts superconductivity in selected bilayers whose constituent materials are nonsuperconducting or only weakly superconducting in the bulk. These results identify quantum-confined bilayers as a promising route to engineering emergent superconductivity in metallic heterostructures.
\end{abstract}

\maketitle

\section{Introduction}

In crystalline superconducting films, such as Pb and Al, quantum confinement has been shown to produce pronounced and often non-monotonic variations of the superconducting critical temperature with thickness, including significant enhancements above the bulk value \cite{nostro}. Confinement effects also modify other electrodynamic properties, such as the magnetic penetration depth and the superfluid density, further demonstrating that superconductivity in ultrathin films can differ substantially from the bulk behavior \cite{forn}. These phenomena originate from confinement-induced modifications of the electronic structure, in particular from changes in the density of states and in the Fermi energy when the film thickness approaches the nanometer scale.

A crucial aspect of quantum confinement in experimentally realized thin films is the presence of unavoidable atomic-scale roughness and structural disorder at interfaces. In contrast to idealized models based on perfectly smooth boundaries, real materials exhibit irregular interfaces that prevent strict quantization of the out-of-plane momentum $k_z$. As a result, the electronic states are not discretized into subbands, but instead undergo a continuous redistribution in momentum space. A recently developed quantum confinement theory \cite{zacbcs,nostro,PhysRevMaterials.9.046001} explicitly incorporates this realistic scenario and provides analytical expressions for thickness-dependent electronic, phononic, and superconducting properties in excellent agreement with experiments, see \cite{Zaccone2025_JPhysMater} for a recent review of this framework. This approach therefore offers a physically grounded description of confinement effects in real heterostructures.

From an intuitive point of view, the confinement mechanism can be understood as a suppression of long-wavelength quasiparticle states along the direction perpendicular to the film. When the film thickness is reduced, quasiparticles with wavelengths larger than the confinement length cannot propagate, which effectively removes a portion of momentum space. This leads to a redistribution of electronic states near the Fermi surface and modifies the density of states at the Fermi level. Since superconductivity in conventional materials depends sensitively on this quantity, even moderate confinement can lead to significant variations of the superconducting critical temperature. Importantly, this mechanism does not rely on idealized subband formation but instead arises naturally from realistic boundary conditions including surface roughness.

In parallel, superconducting proximity effects in multilayer heterostructures allow Cooper pairs to propagate across interfaces, enabling superconducting correlations to develop in adjacent metallic layers. When quantum confinement and proximity effects coexist, they may cooperate and lead to enhanced superconductivity or even to the emergence of superconductivity in systems composed of nonsuperconducting materials.

In this work, we investigate the possibility of constructing two-layer heterostructures in which quantum confinement and proximity effects act simultaneously. We explore bilayer systems where the individual components may be superconducting, weakly superconducting, or even normal metals in the bulk state. By combining confinement-induced modifications of the electronic structure with interlayer proximity coupling, we show that new superconducting states with enhanced critical temperatures can emerge.

To address this problem, we employ the generalized Eliashberg equations including both quantum confinement and proximity effects. In order to apply this approach, three conditions must be satisfied simultaneously: (i) the system must consist of ultrathin layers of two different metals; (ii) the two materials must remain immiscible to avoid alloy formation; and (iii) the Fermi surface can be approximated as spherical, which applies to simple metals such as alkali metals, alkaline earth metals, noble metals, aluminum, and lead. Under these conditions, we identify bilayer combinations that produce superconductivity with critical temperatures higher than those of the individual bulk materials. We emphasize that the model contains no adjustable parameters, since all input quantities are taken from experimentally known material properties of isotropic, one-band, low-$T_c$, $s$-wave superconductors.

This type of problem has also been investigated within different theoretical frameworks and for high-$T_c$ superconductors by other groups \cite{tony}. Here, however, we focus on conventional phonon-mediated superconductivity, described microscopically by Eliashberg theory, in which the pairing interaction is determined by the electron-phonon spectral function $\alpha^2F(\Omega)$ and the residual Coulomb repulsion is accounted for by the Coulomb pseudopotential $\mu^*$ \cite{ummarev,revcarbi}. This framework has proven quantitatively accurate for conventional superconductors and provides a natural starting point for incorporating confinement and proximity effects in bilayer heterostructures.

\section{Summary of the extended Eliashberg framework}

We summarize here the main consequences of quantum confinement within the extended Eliashberg framework. 
In confined metallic systems, the standard Eliashberg theory must be modified because the electronic 
structure becomes thickness dependent. In particular, the \textit{normal density of states} (NDOS), 
the Fermi energy, and the effective interaction parameters entering the Eliashberg equations are all 
renormalized by confinement.

Within this framework:

\begin{itemize}

\item[(i)] The \textit{normal density of states} (NDOS), $N(\varepsilon)$, becomes explicitly energy dependent, reflecting the confinement-induced reconstruction of the electronic spectrum.

\item[(ii)] The Fermi energy is renormalized according to
\begin{equation}
E_F = C(L)^2 E_{F,\mathrm{bulk}},
\end{equation}
where $E_{F,\mathrm{bulk}}$ is the bulk Fermi energy and $C(L)$ is a dimensionless confinement factor depending on the film thickness $L$.

\item[(iii)] The effective electron–phonon coupling constant,
$\lambda$, is modified as
\begin{equation}
\lambda = C(L)\lambda_{\mathrm{bulk}},
\end{equation}
where $\lambda_{\mathrm{bulk}}$ is the electron–phonon coupling constant of the bulk material.

\item[(iv)] The Coulomb pseudopotential $\mu^*$ becomes thickness dependent:
\begin{equation}
\mu^{*}(L)=\frac{C(L)\mu_{\mathrm{bulk}}}{1+\mu_{\mathrm{bulk}}\ln[E_F(L)/\omega_c]},
\end{equation}
where $\mu_{\mathrm{bulk}}$ is the bulk Coulomb pseudopotential and $\omega_c$ is a cutoff energy, typically chosen to be at least three times the maximum phonon energy $\Omega_{\mathrm{max}}$ \cite{Allen}.

\end{itemize}

The dimensionless factor $C(L)$ is fully determined by the confinement geometry and the carrier density, and therefore no adjustable parameters are introduced. This makes the extended Eliashberg approach predictive.

When the system is confined along one spatial direction, as in thin films, the NDOS exhibits two distinct regimes depending on the film thickness $L$ \cite{nostro,zacbcs}.

\subsection{Weak confinement regime: $L>L_c$}

When the thickness is larger than the critical value $L_c$ and the Fermi energy satisfies $E_F>\varepsilon^{*}$, the NDOS takes the form

\begin{equation}
N(\varepsilon)=N(0)C
\left[
{\theta}(\varepsilon^{*}-\varepsilon)\sqrt{\frac{E_{F}}{\varepsilon^{*}}}\frac{|\varepsilon|}{E_{F}}+
{\theta}(\varepsilon-\varepsilon^{*})\sqrt{\frac{|\varepsilon|}{E_{F}}}
\right],
\end{equation}

where $\theta(x)$ is the Heaviside step function, $N(0)$ is the bulk density of states at the Fermi level, and

\begin{align}
C &= \left(1+\frac{1}{3}\frac{L_c^3}{L^3}\right)^{1/3}, \\
\varepsilon^{*} &= \frac{2\pi^{2}\hbar^{2}}{mL^{2}}, \\
L_c &= \left(\frac{2\pi}{n_0}\right)^{1/3}.
\end{align}

Here $m$ is the electron mass, $L$ is the film thickness, and $n_0$ is the carrier density.

In this regime, the Fermi energy and density of states at the Fermi level satisfy

\begin{align}
E_{F} &= C^{2}E_{F,\mathrm{bulk}}, \\
N(E_{F}) &= C N(E_{F,\mathrm{bulk}}) = C N(0),
\end{align}

with

\begin{equation}
N(E_{F,\mathrm{bulk}})=\frac{V(2m)^{3/2}}{2\pi^{2}\hbar^{3}}\sqrt{E_{F,\mathrm{bulk}}}.
\end{equation}

In this regime, the NDOS exhibits a linear dependence on energy for $\varepsilon<\varepsilon^{*}$, while the standard three-dimensional square-root dependence is recovered for $\varepsilon>\varepsilon^{*}$ \cite{nostro,zacbcs}.

\subsection{Strong confinement regime: $L<L_c$}

When $L<L_c$ and $E_F<\varepsilon^{*}$, the NDOS becomes

\begin{equation}
N(\varepsilon)=C'N(0)\sqrt{\frac{E_{F}}{\varepsilon^{*}}}\frac{\varepsilon}{E_{F}},
\end{equation}

where

\begin{align}
E_{F} &= C'^2 E_{F,\mathrm{bulk}}, \\
N(E_{F}) &= C' N(0),
\end{align}

and

\begin{equation}
C'=\frac{2}{\sqrt{L}}\left(\frac{8 \pi}{3}\right)^{2/3}
\frac{1}{\left(n_0(2 \pi)^{3}\right)^{1/6}}.
\end{equation}

In this strong confinement regime, the NDOS is linear in energy:

\begin{equation}
N(\varepsilon)=\sqrt{\frac{E_{F}}{\varepsilon^{*}}}\frac{|\varepsilon|}{E_{F}}.
\end{equation}

Consequently, both the electron–phonon coupling constant and the Coulomb pseudopotential acquire a thickness dependence through the factor $C'$:

\begin{align}
\lambda &= C' \lambda_{\mathrm{bulk}}, \\
\mu^{*} &= \frac{C' \mu_{\mathrm{bulk}}}{1+\mu_{\mathrm{bulk}}\ln(E_{F}/\omega_{c})}.
\end{align}

These confinement-induced modifications of the electronic structure constitute the key ingredients of the extended Eliashberg framework used in the following sections.

\begin{figure*}[t!]
	\centerline{\includegraphics[width=1\textwidth]{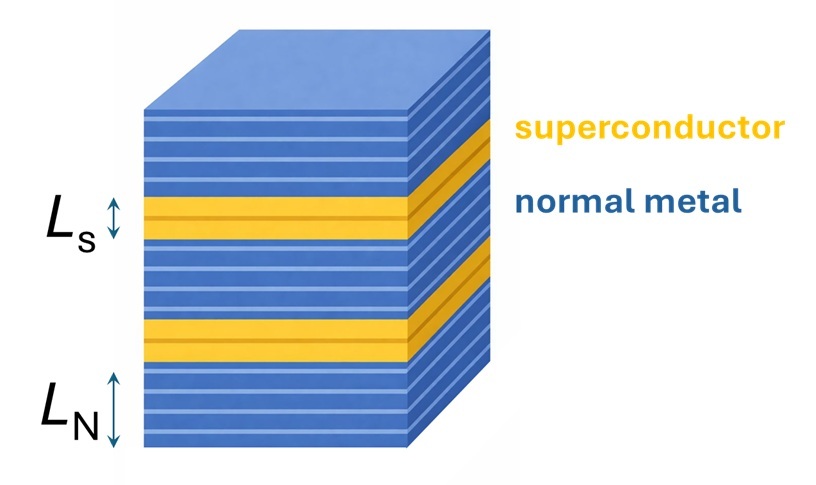}}
\caption{Schematic illustration of a superconductor/normal-metal multilayer system combining quantum confinement and the superconducting proximity effect.
The superconducting layer ($S$) and the normal-metal layer ($N$) have thicknesses $L_S$ and $L_N$, respectively.
Quantum confinement arises from the finite thickness of each layer, while interlayer coupling enables the proximity-induced transfer of superconducting correlations across the interface.}
\label{fig:bilayer_scheme}
\end{figure*}

\section{Eliashberg theory with confinement and proximity effects}

We now extend the quantum-confinement framework by incorporating superconducting
proximity effects in layered heterostructures. Instead of considering isolated ultrathin
films—whose fabrication becomes increasingly challenging at sub-nanometer
thickness—we investigate multilayer structures composed of alternating superconducting
and normal-metal layers. Such heterostructures can be reliably fabricated using sputtering,
molecular beam epitaxy, or related thin-film deposition techniques. In these systems,
quantum confinement naturally arises from the reduced thickness of each layer, while
superconducting correlations propagate across the interfaces via the proximity effect. 
A schematic representation of the layered geometry and of the interplay between quantum
confinement and proximity coupling is shown in Fig.~\ref{fig:bilayer_scheme}.

To describe superconductivity in the presence of both confinement and proximity effects,
we employ the isotropic one-band $s$-wave Eliashberg theory generalized to multilayer
systems. In this case, four coupled equations must be solved self-consistently for the
superconducting gaps $\Delta_{S(N)}(i\omega_n)$ and the mass-renormalization functions
$Z_{S(N)}(i\omega_n)$ in the superconducting ($S$) and normal ($N$) layers. 
Here $\omega_n=\pi T(2n+1)$ are the fermionic Matsubara frequencies, and the functions
$Z(i\omega_n)$ account for the renormalization of the quasiparticle mass due to
electron–phonon interactions.

Within the Migdal approximation \cite{Mig}, the Eliashberg equations on the imaginary
axis read \cite{revnostro,McMillan,Ummaprox1,Ummaprox2,Ummaprox3,Carbi1,Carbi2,Carbi3,Carbi4,kresin}:

\begin{eqnarray}
\omega_{n}Z_{N}(i\omega_{n})=\omega_{n}+ \pi T\sum_{m}\Lambda^{Z}_{N}(i\omega_{n},i\omega_{m})N^{Z}_{N}(i\omega_{m})+\Gamma_{N} N^{Z}_{S}(i\omega_{n})
\label{eq:EE1}
\end{eqnarray}

\begin{eqnarray}
&&Z_{N}(i\omega_{n})\Delta_{N}(i\omega_{n})=\pi
T\sum_{m}\big[\Lambda^{\Delta}_{N}(i\omega_{n},i\omega_{m})-\mu^{*}_{N}(\omega_{c})\big]\times\nonumber\\
&&\times\Theta(\omega_{c}-|\omega_{m}|)N^{\Delta}_{N}(i\omega_{m})
+\Gamma_{N} N^{\Delta}_{S}(i\omega_{n})
 \label{eq:EE2}
\end{eqnarray}

\begin{eqnarray}
\omega_{n}Z_{S}(i\omega_{n})=\omega_{n}+ \pi T\sum_{m}\Lambda^{Z}_{S}(i\omega_{n},i\omega_{m})N^{Z}_{S}(i\omega_{m})+\Gamma_{S} N^{Z}_{N}(i\omega_{n})
\label{eq:EE3}
\end{eqnarray}

\begin{eqnarray}
&&Z_{S}(i\omega_{n})\Delta_{S}(i\omega_{n})=\pi
T\sum_{m}\big[\Lambda^{\Delta}_{S}(i\omega_{n},i\omega_{m})-\mu^{*}_{S}(\omega_{c})\big]\times\nonumber\\
&&\times\Theta(\omega_{c}-|\omega_{m}|)N^{\Delta}_{S}(i\omega_{m})
+\Gamma_{S}N^{\Delta}_{N}(i\omega_{n})
 \label{eq:EE4}
\end{eqnarray}

where the functions $N^{\Delta}_{S(N)}$ and $N^{Z}_{S(N)}$ describe the quasiparticle
density contributions entering the self-consistent equations:

\begin{eqnarray}
&&N^{\Delta}_{S(N)}(i\omega_{m})=
\frac{\Delta_{S(N)}(i\omega_{m})}
{\sqrt{\omega^{2}_{m}+\Delta^{2}_{S(N)}(i\omega_{m})}}
\left[\frac{N_{S(N)}(i\omega_{m})+N_{S(N)}(-i\omega_{m})}{2}\right]\times\nonumber\\
&&\times\frac{2}{\pi} \arctan\!\left(\frac{W_{S(N)}}{2Z_{S(N)}(i\omega_{m})\sqrt{\omega_{m}^{2}+\Delta_{S(N)}^{2}(i\omega_{m})}}\right),
\end{eqnarray}

\begin{eqnarray}
&&N^{Z}_{S(N)}(i\omega_{m})=
\frac{\omega_{m}}
{\sqrt{\omega^{2}_{m}+\Delta^{2}_{S(N)}(i\omega_{m})}}
\left[\frac{N_{S(N)}(i\omega_{m})+N_{S(N)}(-i\omega_{m})}{2}\right]\times\nonumber\\
&&\times\frac{2}{\pi} \arctan\!\left(\frac{W_{S(N)}}{2Z_{S(N)}(i\omega_{m})\sqrt{\omega_{m}^{2}+\Delta_{S(N)}^{2}(i\omega_{m})}}\right).
\end{eqnarray}

The pairing kernels $\Lambda_{S(N)}$ are defined as

\begin{equation}
\Lambda_{S(N)}(i\omega_{n},i\omega_{m})=
2\int_{0}^{+\infty}d\Omega\,
\frac{\Omega\,\alpha^{2}_{S(N)}F(\Omega)}
{(\omega_{n}-\omega_{m})^{2}+\Omega^{2}},
\end{equation}

where $\alpha^{2}_{S(N)}F(\Omega)$ are the electron–phonon spectral functions of the
superconducting and normal layers. The corresponding electron–phonon coupling
constants are given by

\begin{equation}
\lambda_{S(N)}=
2\int_{0}^{+\infty}d\Omega\,
\frac{\alpha^{2}_{S(N)}F(\Omega)}{\Omega}.
\end{equation}

The coupling between adjacent layers is described by the tunneling parameters

\begin{equation}
\Gamma_{S(N)}=\pi |t|^{2} A L_{N(S)} N_{N(S)}(0),
\label{eq:EE6}
\end{equation}

where $|t|^{2}$ is the transmission probability across the interface, $A$ is the
junction cross-sectional area, $L_{S(N)}$ are the layer thicknesses, and
$N_{S(N)}(0)$ are the densities of states at the Fermi level. The bandwidths
$W_{S(N)}$ are taken as half of the corresponding Fermi energies.

As in the single-layer case, quantum confinement modifies the material parameters in
each layer:

\begin{equation}
\lambda_{S(N)} = C_{S(N)}\,\lambda_{S(N),\mathrm{bulk}},
\end{equation}

\begin{equation}
\mu^{*}_{S(N)}=
\frac{C_{S(N)}\,\mu_{S(N),\mathrm{bulk}}}
{1+\mu_{S(N),\mathrm{bulk}}\ln(E_{F,S(N)}/\omega_c)},
\end{equation}

\begin{equation}
E_{F,S(N)}=C_{S(N)}^{2}E_{F,S(N),\mathrm{bulk}},
\end{equation}

\begin{equation}
N(E_{F,S(N)})=C_{S(N)}\,N_{S(N),\mathrm{bulk}}(0).
\end{equation}

To solve the coupled equations, nine input parameters are required: the two electron–phonon
spectral functions $\alpha^{2}_{S(N)}F(\Omega)$, the two Coulomb pseudopotentials
$\mu^{*}_{S(N)}(\omega_c)$, the two densities of states $N_{S(N)}(0)$, the two layer
thicknesses $L_{S(N)}$, and the product $A|t|^{2}$. The geometrical parameters are taken
from experimental values. We assume ideal interfaces with $|t|^{2}=1$ and take
$A=10^{-7}$~m$^{2}$; the final results are independent of $A$.

The Eliashberg equations are solved numerically assuming equal layer thicknesses,
although the formalism allows for arbitrary thickness ratios. The input parameters
required for the calculations are summarized in Table~\ref{tab:params}. The electron–phonon
spectral functions are taken from previous works \cite{nostro,noblemetal,Mg,revnostro}.
We emphasize that the present theory contains no adjustable parameters.
\begin{table}[t]
\centering
\small
\setlength{\tabcolsep}{3.5pt}
\renewcommand{\arraystretch}{1.15}
\resizebox{\textwidth}{!}{%
\begin{tabular}{|c|c|c|c|c|c|c|c|c|c|}
  \hline
  metal & $\lambda_{bulk}$ & $\mu^{*}(\omega_c)$ & $\omega_c$ (meV) & $n_0$ ($10^{28}$ m$^{-3}$) & $E_{F,bulk}$ (eV) & $T_{c,bulk}$ (K) & $L_c$ (\AA) & $\Omega_{max}$ (meV)& $N(E_F)$ ($eV^{-1}$) \\
  \hline
  Ag & 0.160 & 0.11000 & 75 & 5.86 & 5.49  & 0.000 & 4.75& 80 & 0.13000\\
  Al & 0.430 & 0.13984 & 145& 18.10& 11.70 & 1.200 & 3.26&150 & 0.17570\\
  Au & 0.220 & 0.11000 & 55 & 5.90 & 5.54  & 0.000 & 4.74&60  & 0.14000\\
  Be & 0.210 & 0.07700 & 255& 14.60& 24.70 & 0.026 & 2.94&260 & 0.03235\\
  Cs & 0.180 & 0.14000 & 90 & 0.91 & 1.59  & 0.000 & 8.84&100 & 0.01960\\
  Na & 0.180 & 0.15000 & 46 & 2.65 & 3.24  & 0.000 & 6.19&30  & 0.22350\\
  Mg & 0.300 & 0.16000 & 90 & 8.61 & 7.08  & 0.000 & 4.18&100 & 0.23380\\
  Rb & 0.065 & 0.12600 & 21 & 1.15 & 1.99  & 0.000 & 8.18&22  & 0.47350\\
  Pb & 1.550 & 0.13202 & 75 & 9.47 & 13.2  & 7.220 & 3.62&80  & 0.25866\\
  \hline
\end{tabular}%

}
\caption{Input bulk parameters, bulk critical temperature and critical thickness. The last input parameter present in the theory, besides these, is the film tickness $L$ and the electron-phonon spectral functions $\alpha^{2}F(\Omega)$. The reference for these data are in our previous papers \cite{nostro,noblemetal,Mg,revnostro}}
\label{tab:params}
\end{table}
\subsection{Bilayer Al/Mg}
\begin{figure*}[t!]	\centerline{\includegraphics[width=1\textwidth]{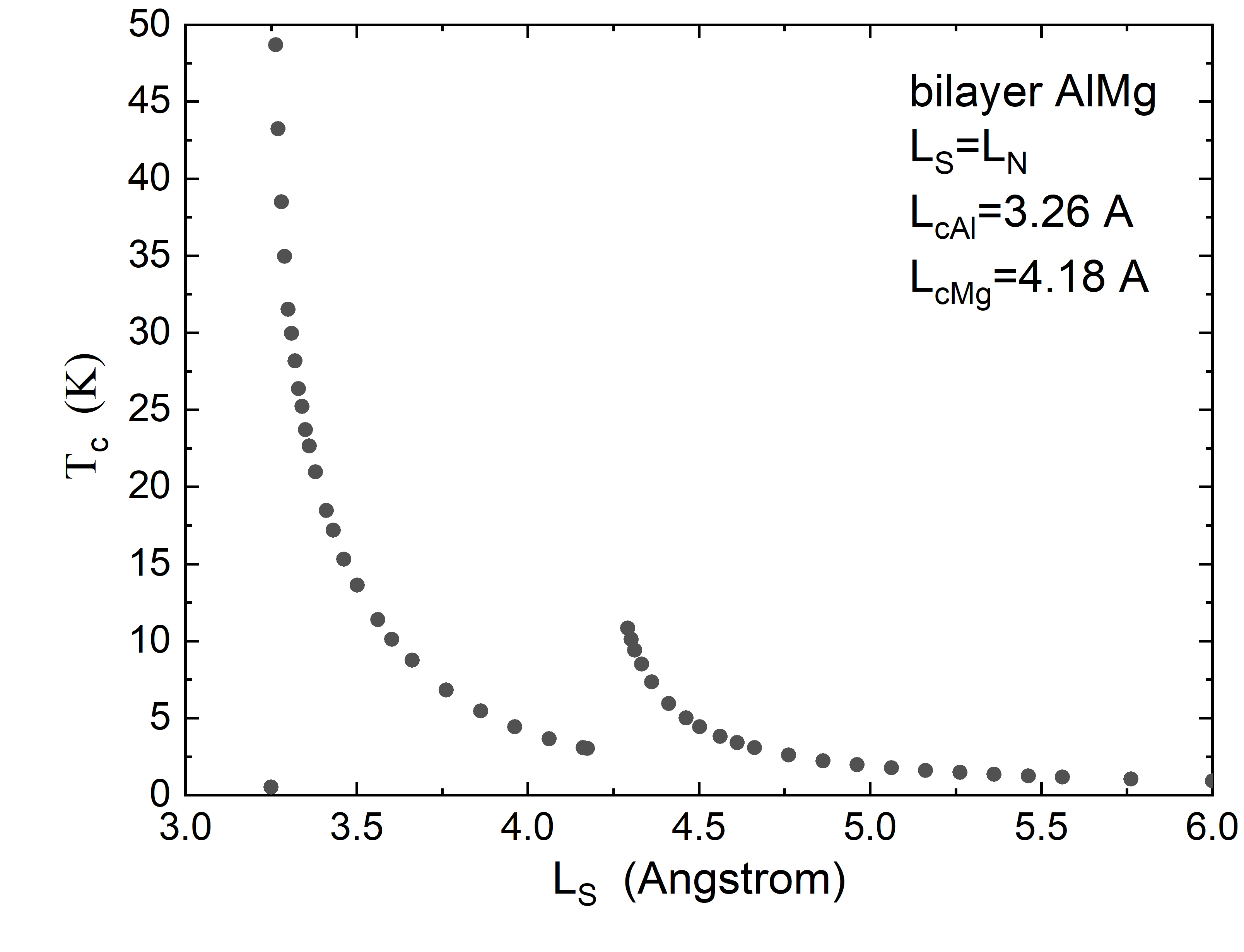}}
\caption{Critical temperature $T_c$ as a function of the single-layer thickness $L_S$ for an $Al/Mg$ superconductor $Al$-normal metal $Mg$ bilayer with equal layer thicknesses ($L_S=L_N$). Full symbols denote numerical solutions of the confinement- and proximity-modified Eliashberg equations.}
\label{fig:bilayer_tc1}
\end{figure*}

 We consider as first case a superconductor/normal-metal bilayer composed of aluminum(Al)and magnesium(Mg). This choice is motivated by two considerations. First, Al and Mg do not form stable alloys under the relevant growth conditions, ensuring that the layers remain chemically distinct, in contrast to combinations such as Au/Ag. Second, both materials are electronically simple, and all parameters required by the theory are well established experimentally or from \textit{ab initio} calculations. For simplicity, we assume equal thicknesses for the superconducting and normal layers, $L_S=L_N$, although the formalism readily allows for unequal thicknesses.
The resulting critical temperature as a function of thickness is shown in Fig.~\ref{fig:bilayer_tc1}. Two key features emerge: first, due to quantum confinement, the critical temperature exceeds the bulk aluminum value in selected thickness ranges; second, the dependence of $T_c$ on thickness is strongly non-monotonic. This behavior arises from the mismatch between the critical confinement lengths $L_{cS}$ and $L_{cN}$, leading to a discontinuity in $T_c(L)$ at $L=L_{cN}$.

Specifically, Fig.~\ref{fig:bilayer_tc1} shows that the critical temperature $T_c$ depends non-monotonically on the layer thickness: at selected thicknesses, quantum confinement independently enhances superconducting pairing in the Mg layer and, simultaneously, strengthens proximity coupling with the Al layer, leading to pronounced maxima in $T_c$. The discontinuity in $T_c$ signals the thickness at which the confinement regimes of the two materials change relative to one another, abruptly modifying the balance between intrinsic confinement-induced superconductivity in Mg and proximity-induced pairing from Al.

\begin{figure*}[t!]	\centerline{\includegraphics[width=1\textwidth]{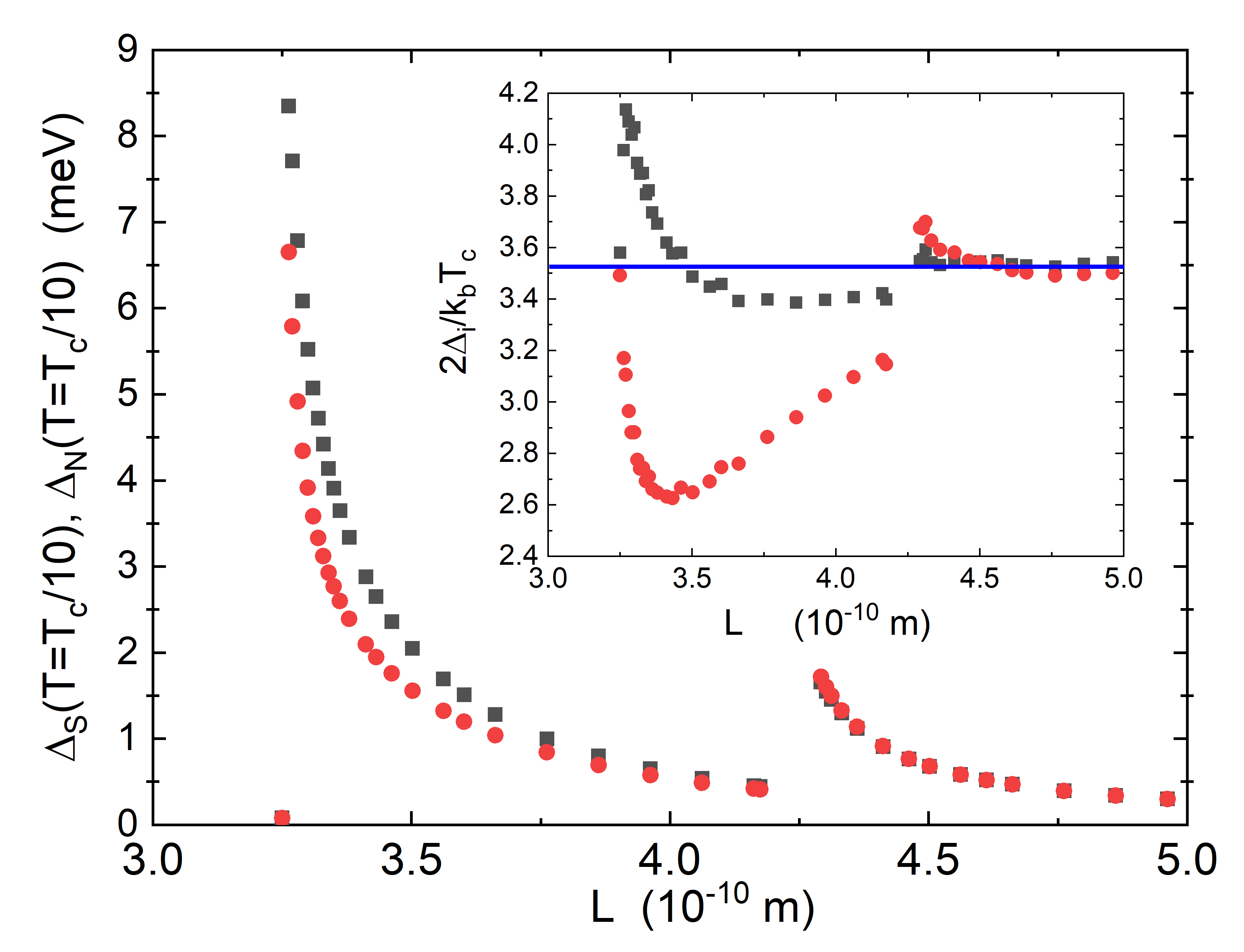}}
\caption{Low temperature gaps values in the superconductive $\Delta_S$ (black full squares) and normal (red full circles) $\Delta_N$ layer as a function of the single-layer thickness $L_S$ for an $Al/Mg$ superconductor $Al$-normal metal $Mg$ bilayer with equal layer thicknesses ($L_S=L_N$). Full symbols denote numerical solutions of the confinement- and proximity-modified Eliashberg equations.}
\label{fig:bilayer_delta2}
\end{figure*}

The Fig.~\ref{fig:bilayer_delta2}
shows the dependence of the values of the two low temperature gaps on the thickness of the layers. Here, the values of the two gaps are similar due to our particular choice of $L_S=L_N$, but they can also be very different.
The inset shows the value of the $2\Delta_{S,N}/k_BT_c$ ratio as a function of thickness. As can be seen, for some values of $L$, this ratio is smaller than the BCS value, which is completely impossible in traditional BCS and Eliashberg theories without quantum confinement and the proximity effect.

\subsection{Bilayer Pb/Mg}

\begin{figure*}[t!]
\centerline{\includegraphics[width=1\textwidth]{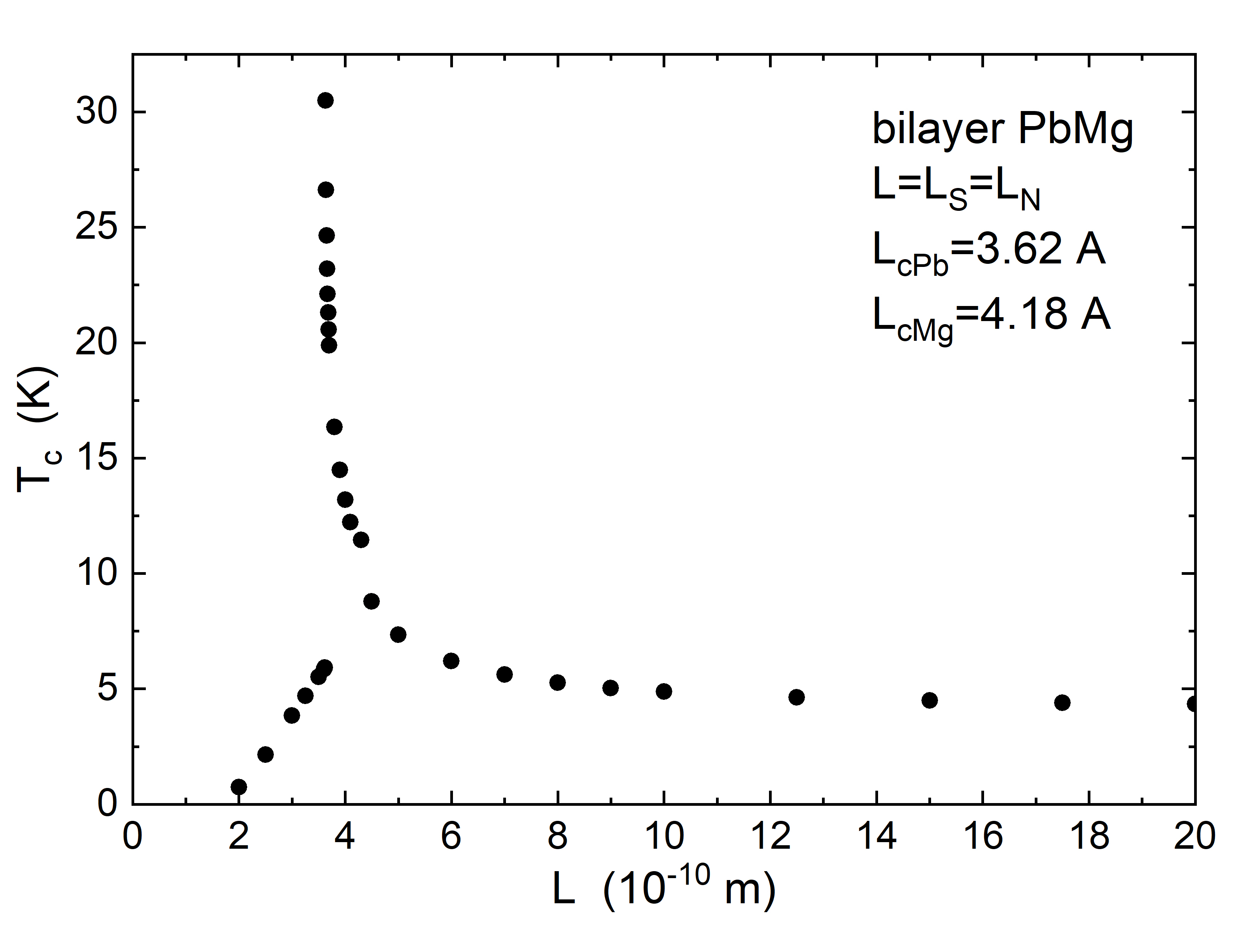}}
\caption{Critical temperature $T_c$ as a function of the single-layer thickness $L_S$ for a Pb/Mg bilayer with equal layer thicknesses ($L_S=L_N$). Full symbols denote numerical solutions of the confinement- and proximity-modified Eliashberg equations.}
\label{fig:bilayer_tc3}
\end{figure*}

We first consider a superconductor/normal-metal bilayer composed of lead (Pb) and magnesium (Mg), shown in Fig.~\ref{fig:bilayer_tc3}. In this configuration, Pb acts as the superconducting layer, while Mg is a normal metal in the bulk. However, due to quantum confinement, Mg can develop superconducting correlations at sufficiently small thicknesses, which enhances the proximity coupling between the two layers.

The calculated critical temperature $T_c$ exhibits a non-monotonic dependence on the layer thickness. Small discontinuities appear near the critical thickness values $L_S$ and $L_N$, corresponding to changes in the confinement regime. Since these critical thicknesses are very close to each other for Pb and Mg, the resulting features appear as weak kinks rather than pronounced jumps. Overall, the interplay between confinement and proximity effects leads to an enhancement of $T_c$ relative to the bulk Pb/Mg bilayer.

\subsection{Bilayer Pb/Ag}

\begin{figure*}[t!]
\centerline{\includegraphics[width=1\textwidth]{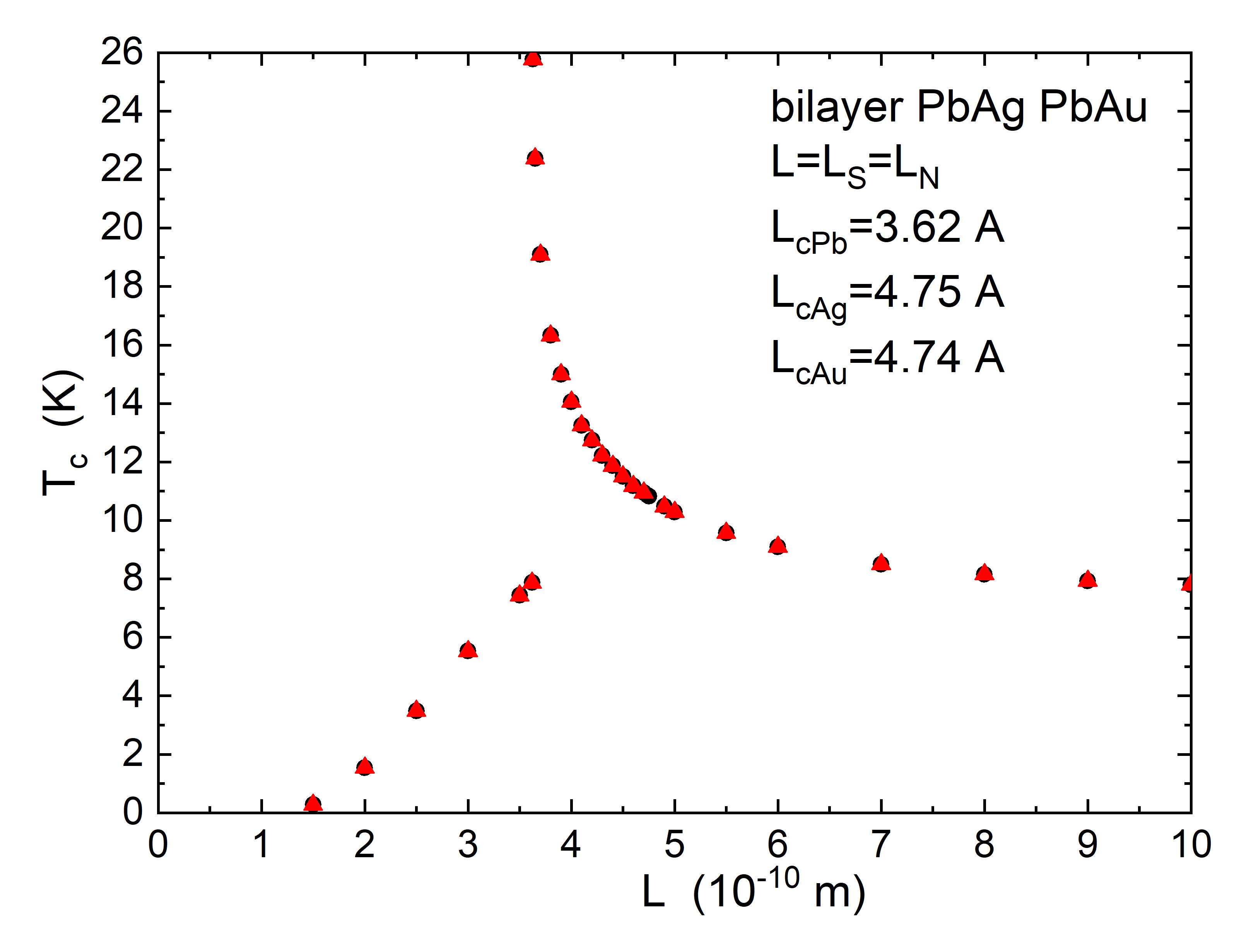}}
\caption{Critical temperature $T_c$ as a function of the single-layer thickness $L_S$ for a Pb/Ag bilayer with equal layer thicknesses ($L_S=L_N$). Full symbols denote numerical solutions of the confinement- and proximity-modified Eliashberg equations.}
\label{fig:bilayer_tc4}
\end{figure*}

We next consider a Pb/Ag bilayer, shown in Fig.~\ref{fig:bilayer_tc4}, where lead acts as the superconducting layer and silver as the normal metal. In this case, the proximity effect induces superconducting correlations in the Ag layer, while quantum confinement modifies the electronic properties of both layers.

The resulting critical temperature exhibits a smooth thickness dependence, with modest enhancement at small thicknesses. We also performed analogous calculations replacing silver with gold, obtaining nearly identical results. This similarity arises from the comparable electronic parameters of Ag and Au, including carrier density, Fermi energy, and electron-phonon coupling, which enter the Eliashberg framework.

\subsection{Bilayer Be/Mg}

\begin{figure*}[t!]
\centerline{\includegraphics[width=1\textwidth]{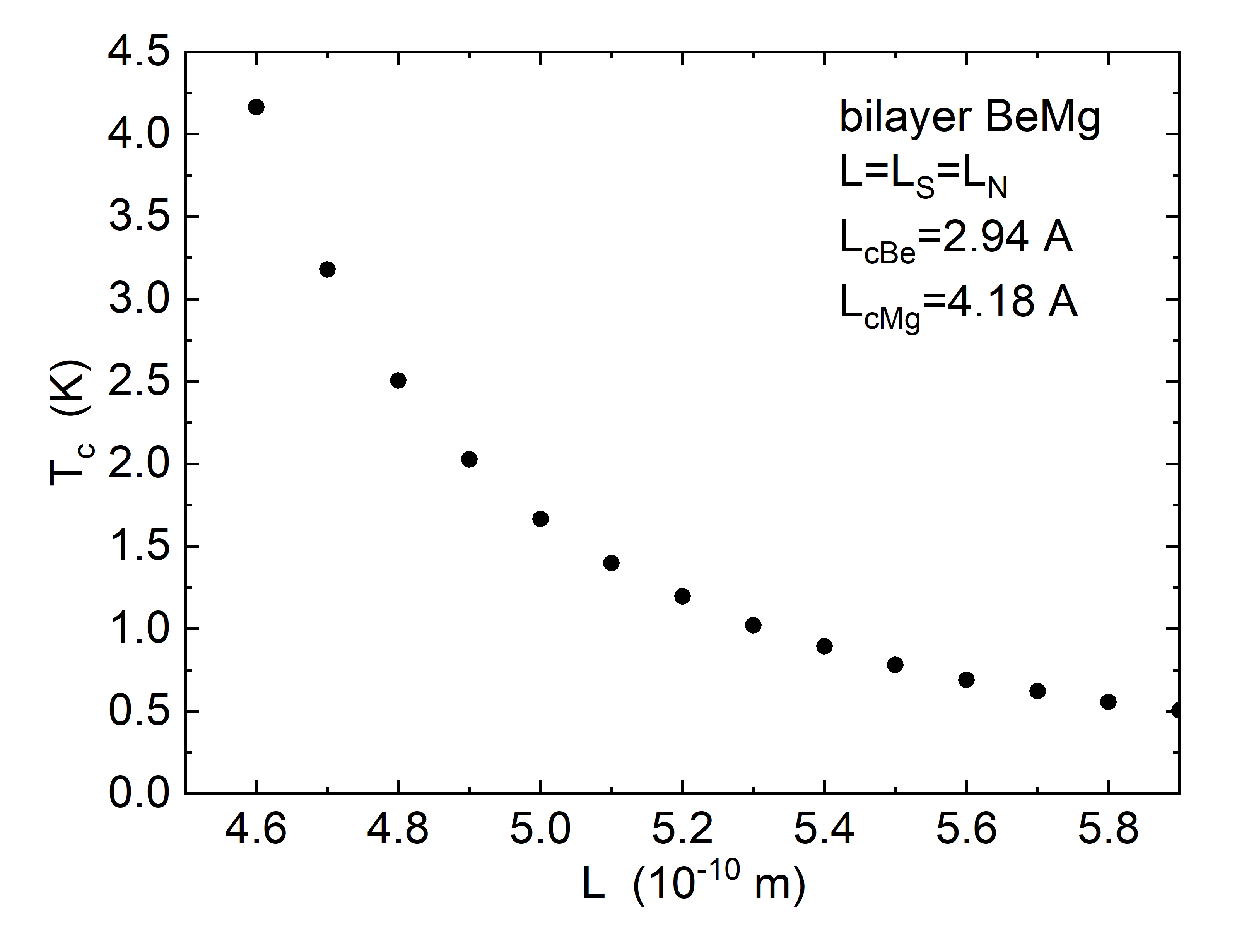}}
\caption{Critical temperature $T_c$ as a function of the single-layer thickness $L_S$ for a Be/Mg bilayer with equal layer thicknesses ($L_S=L_N$). Full symbols denote numerical solutions of the confinement- and proximity-modified Eliashberg equations.}
\label{fig:bilayer_tc5}
\end{figure*}

We now consider a bilayer composed of beryllium (Be) and magnesium (Mg), shown in Fig.~\ref{fig:bilayer_tc5}. In this system, Be acts as a weak superconductor, while Mg is a normal metal in the bulk. Quantum confinement enhances the superconducting properties of Be and may induce superconductivity in Mg at very small thicknesses.

The resulting $T_c$ behavior reflects the combined effects of confinement-induced enhancement and proximity coupling. As the layer thickness decreases, the superconducting correlations become stronger, leading to a noticeable increase in $T_c$ at small thicknesses.

\subsection{Bilayer Mg/Rb}

\begin{figure*}[t!]
\centerline{\includegraphics[width=1\textwidth]{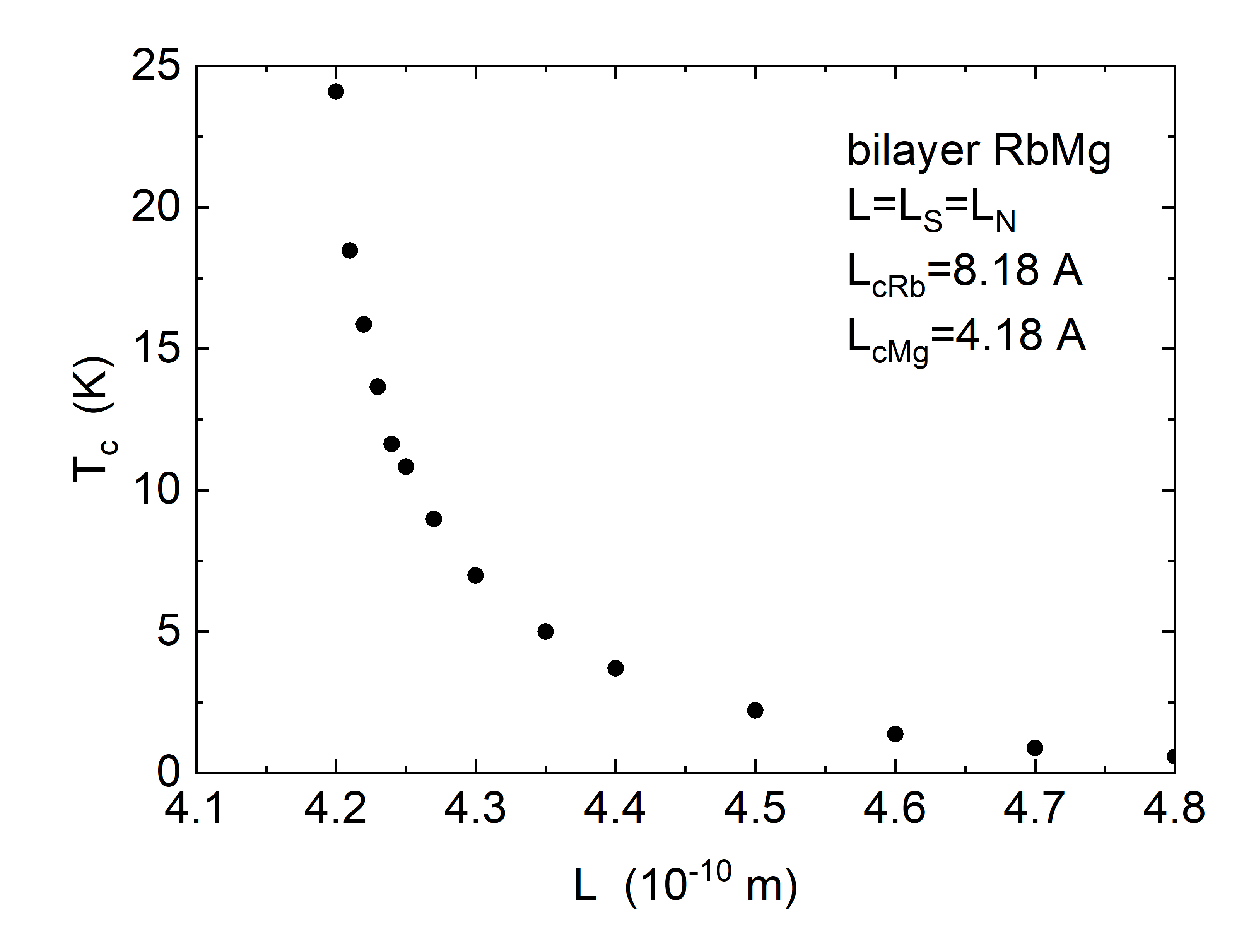}}
\caption{Critical temperature $T_c$ as a function of the single-layer thickness $L_S$ for an Mg/Rb bilayer with equal layer thicknesses ($L_S=L_N$). Full symbols denote numerical solutions of the confinement- and proximity-modified Eliashberg equations.}
\label{fig:bilayer_tc6}
\end{figure*}

We now turn to a particularly interesting case consisting of two normal metals, magnesium (Mg) and rubidium (Rb), shown in Fig.~\ref{fig:bilayer_tc6}. In the bulk, neither material is superconducting. However, when Mg is sufficiently confined, it becomes superconducting due to the enhancement of the density of states at the Fermi level.

This confinement-induced superconductivity in Mg enables proximity coupling with the Rb layer. As a result, the bilayer exhibits a finite critical temperature for sufficiently small thicknesses. Unlike previous cases, however, the superconductivity disappears beyond a critical thickness, and $T_c$ drops to zero when confinement effects become too weak.

\subsection{Bilayer Mg/Na}

\begin{figure*}[t!]
\centerline{\includegraphics[width=1\textwidth]{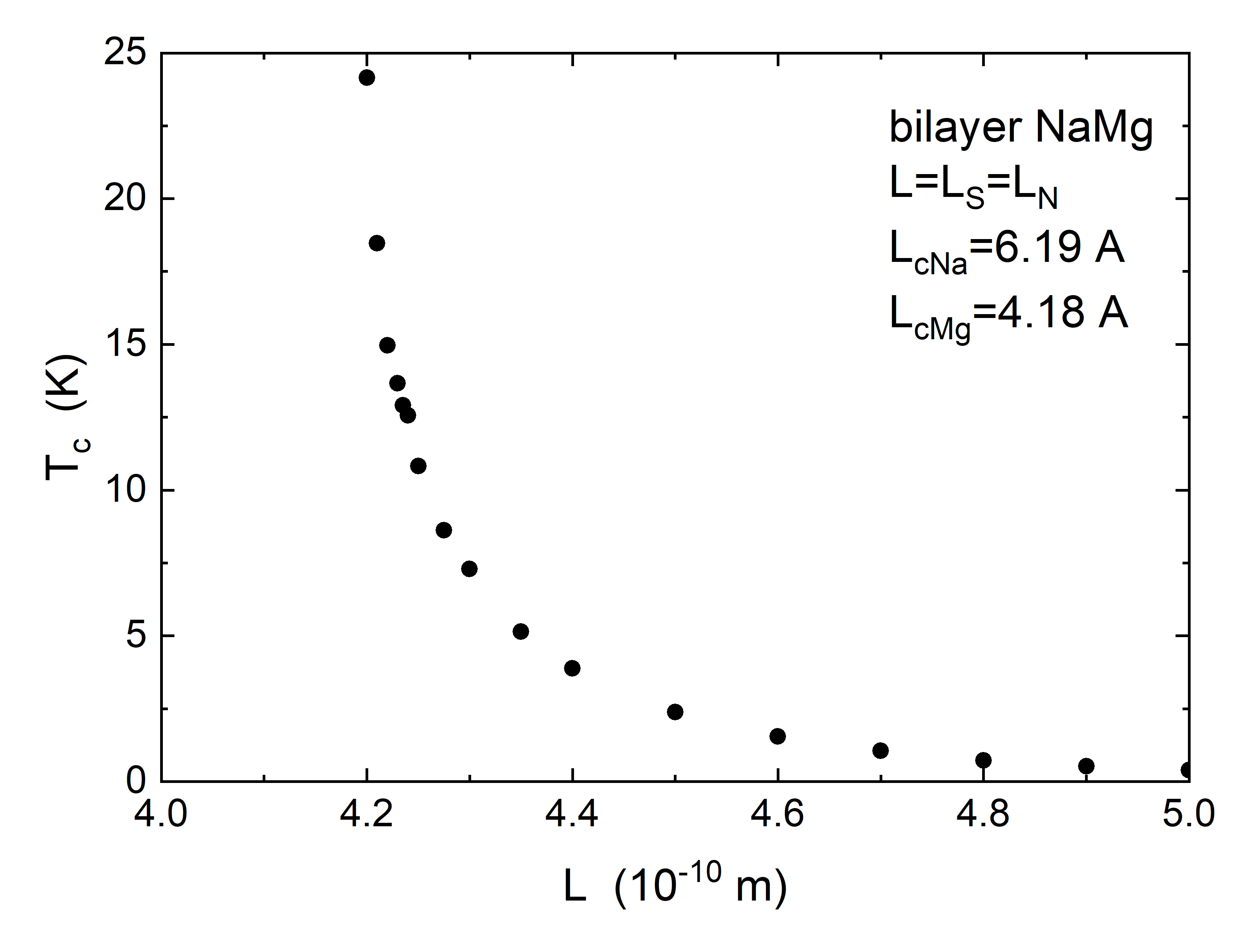}}
\caption{Critical temperature $T_c$ as a function of the single-layer thickness $L_S$ for an Mg/Na bilayer with equal layer thicknesses ($L_S=L_N$). Full symbols denote numerical solutions of the confinement- and proximity-modified Eliashberg equations.}
\label{fig:bilayer_tc7}
\end{figure*}

A similar situation occurs for the Mg/Na bilayer shown in Fig.~\ref{fig:bilayer_tc7}. Both magnesium and sodium are normal metals in the bulk. However, quantum confinement induces superconductivity in Mg at sufficiently small thicknesses, allowing proximity coupling to the Na layer.

As in the Mg/Rb case, superconductivity emerges only in the strong confinement regime. The critical temperature decreases rapidly as the layer thickness increases, eventually vanishing when confinement effects become negligible.

\subsection{Bilayer Mg/Cs}

\begin{figure*}[t!]
\centerline{\includegraphics[width=1\textwidth]{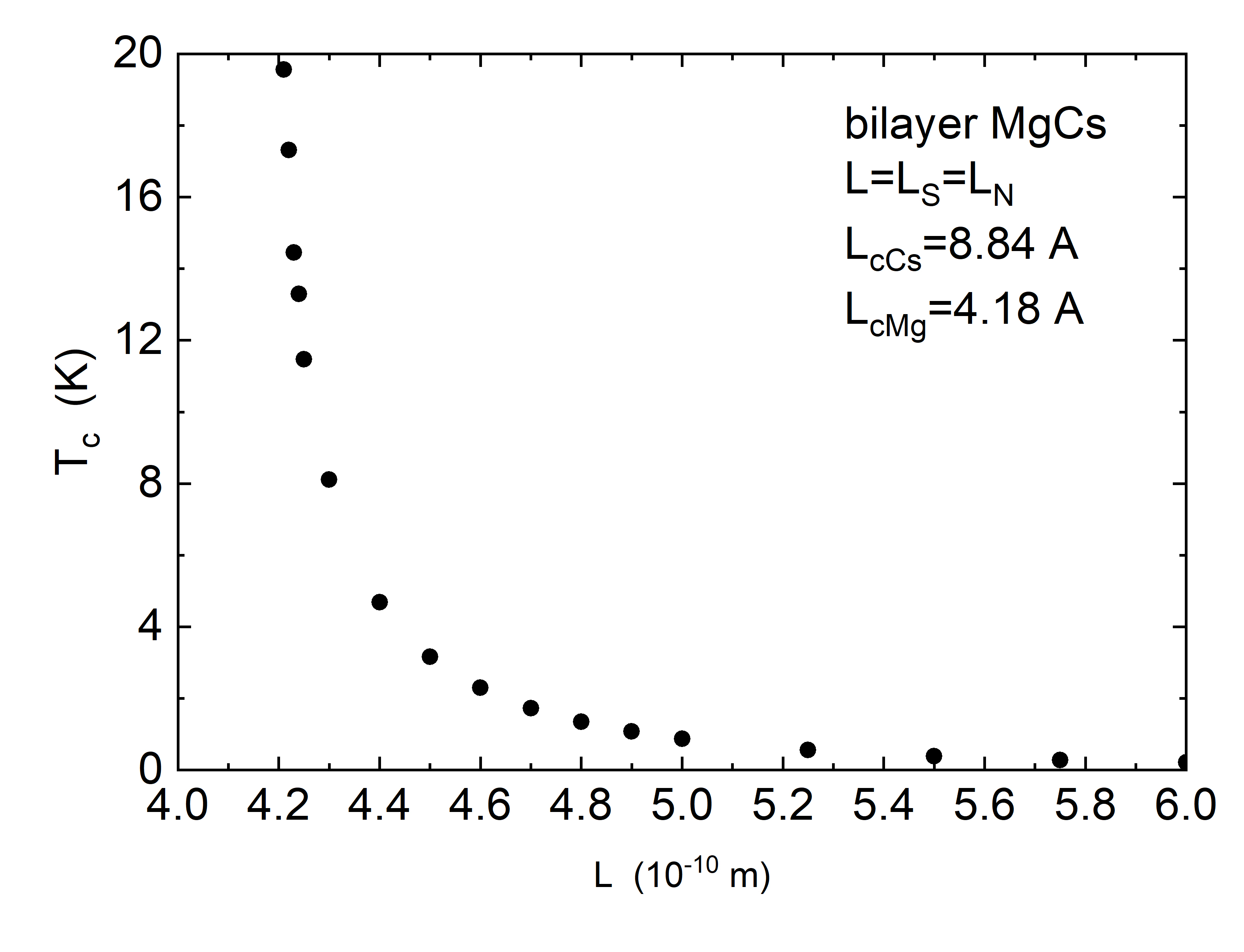}}
\caption{Critical temperature $T_c$ as a function of the single-layer thickness $L_S$ for an Mg/Cs bilayer with equal layer thicknesses ($L_S=L_N$). Full symbols denote numerical solutions of the confinement- and proximity-modified Eliashberg equations.}
\label{fig:bilayer_tc8}
\end{figure*}

Another example involving two normal metals is the Mg/Cs bilayer, shown in Fig.~\ref{fig:bilayer_tc8}. As in the previous cases, confinement induces superconductivity in Mg at small thicknesses, which in turn leads to proximity-induced superconducting correlations in Cs.

The resulting $T_c$ shows a strong dependence on thickness, with superconductivity appearing only in the ultrathin regime and disappearing at larger thicknesses.

\subsection{Bilayer Pb/Al}

\begin{figure*}[t!]
\centerline{\includegraphics[width=1\textwidth]{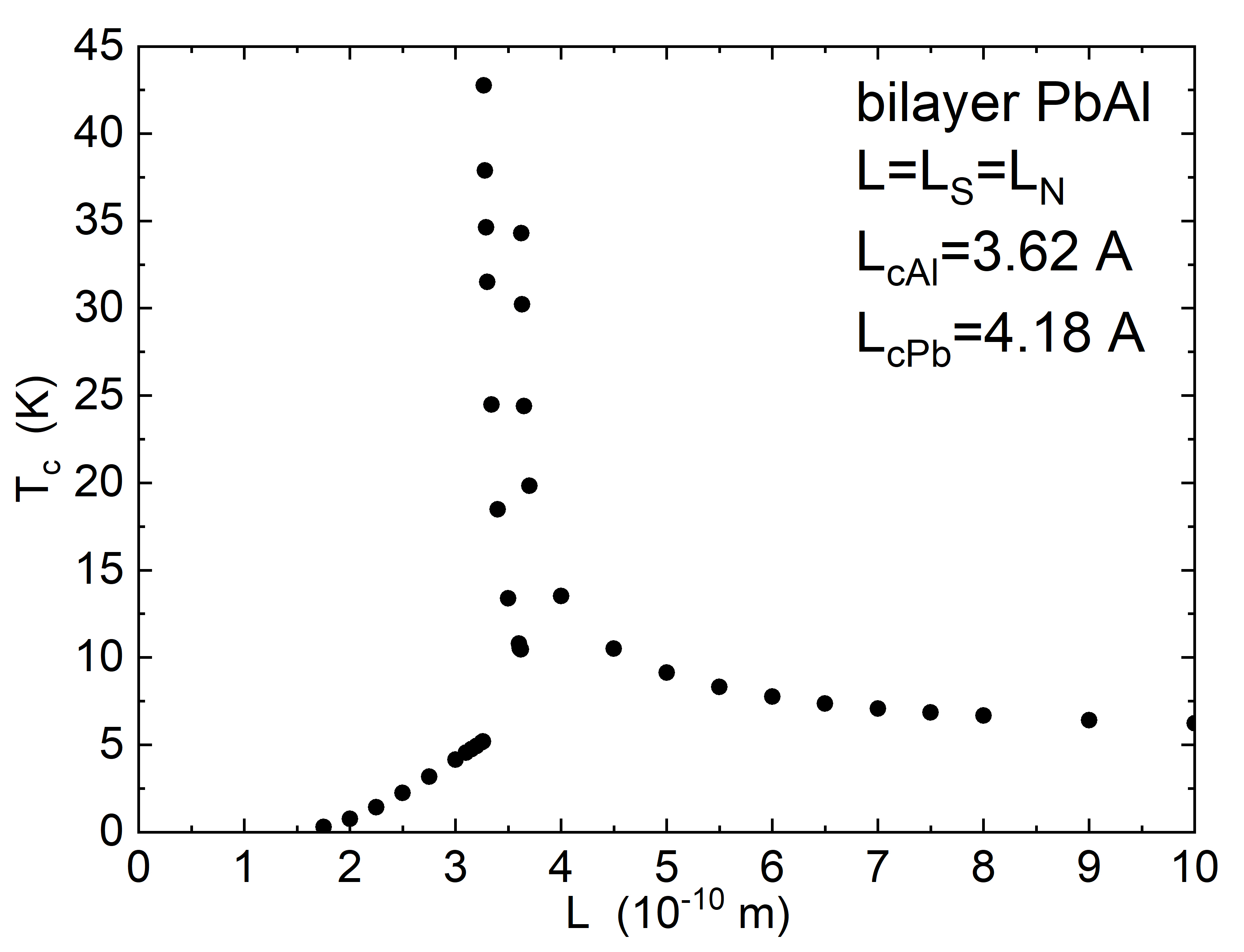}}
\caption{Critical temperature $T_c$ as a function of the single-layer thickness $L_S$ for a Pb/Al bilayer with equal layer thicknesses ($L_S=L_N$). Full symbols denote numerical solutions of the confinement- and proximity-modified Eliashberg equations.}
\label{fig:bilayer_tc9}
\end{figure*}

We now consider a superconductor/superconductor bilayer composed of lead (Pb) and aluminum (Al), shown in Fig.~\ref{fig:bilayer_tc9}. Although both materials are superconducting in the bulk, their critical temperatures differ significantly ($T_{c,Pb}=7.22$~K and $T_{c,Al}=1.2$~K). 

Between these two temperatures, aluminum behaves effectively as a normal metal and is driven superconducting by proximity to Pb. Quantum confinement further enhances the superconducting properties, leading to a modified $T_c$ compared with bulk values.

\subsection{Bilayer Pb/Be}

\begin{figure*}[t!]
\centerline{\includegraphics[width=1\textwidth]{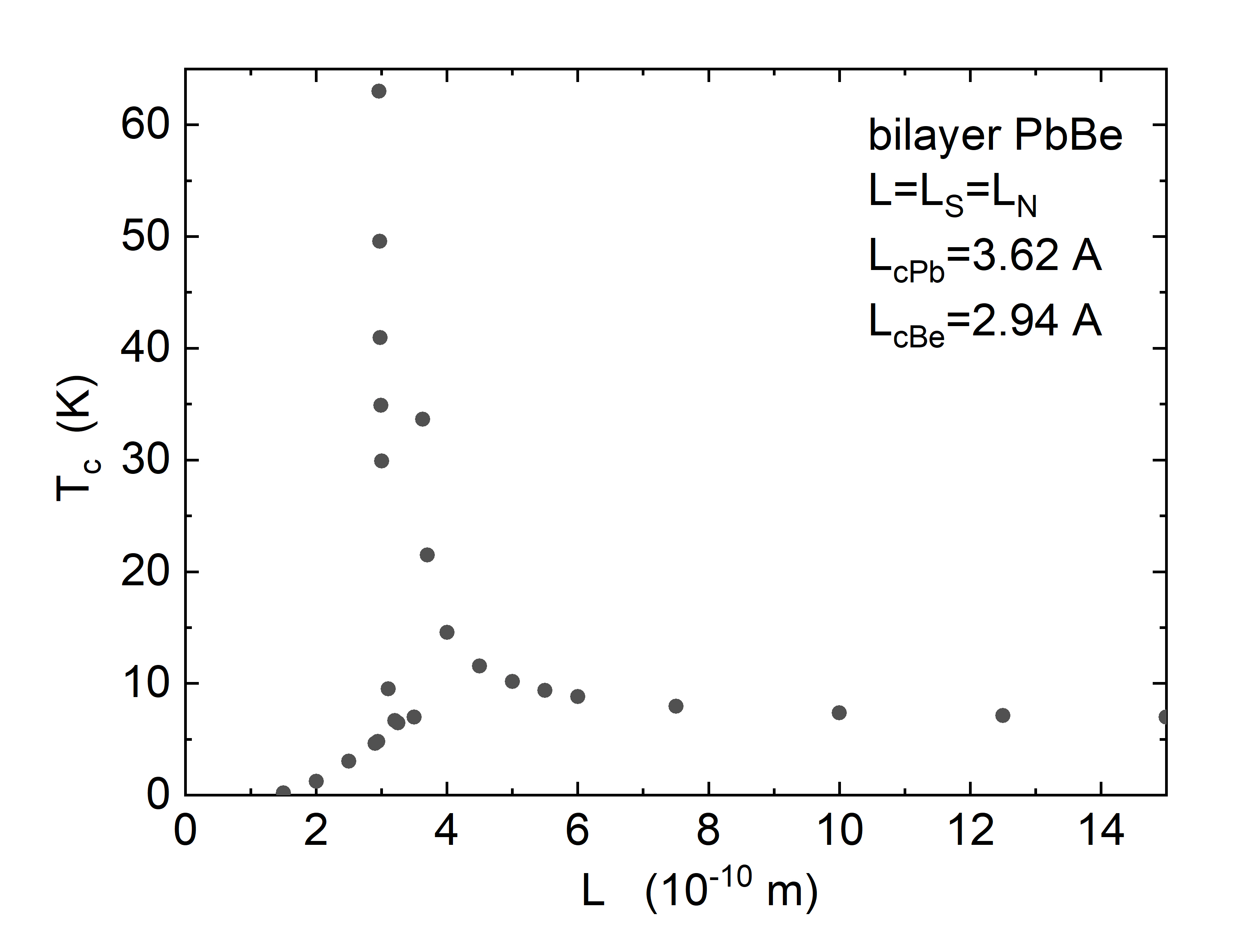}}
\caption{Critical temperature $T_c$ as a function of the single-layer thickness $L_S$ for a Pb/Be bilayer with equal layer thicknesses ($L_S=L_N$). Full symbols denote numerical solutions of the confinement- and proximity-modified Eliashberg equations.}
\label{fig:bilayer_tc10}
\end{figure*}

Finally, we consider a Pb/Be bilayer, shown in Fig.~\ref{fig:bilayer_tc10}. Lead is a strong superconductor, while beryllium is a very weak superconductor with $T_{c,Be}=0.026$~K. 

In this system, proximity effects and confinement enhance superconductivity in the Be layer, leading to an overall increase in $T_c$ at small thicknesses. As in previous cases, the interplay between confinement and proximity effects governs the thickness dependence of the critical temperature.

These results demonstrate that nanoscale heterostructuring provides a powerful route to engineer superconducting properties beyond bulk material limitations. By combining quantum confinement and proximity effects, superconductivity can be enhanced, tuned, or even induced in systems that are non-superconducting in bulk form.
\section{Conclusions}

In this work, we have shown that combining quantum confinement with the superconducting proximity effect in layered heterostructures provides a robust and experimentally realistic route to engineering superconductivity at the nanoscale. By considering bilayer systems composed of ultrathin superconducting and normal-metal layers, we demonstrated that confinement-induced modifications of the electronic structure can cooperate with interlayer coupling to enhance superconducting properties.

In particular, we found that, within specific thickness ranges, the superconducting critical temperature can exceed that of the bulk superconducting constituent. This enhancement originates from the confinement-induced redistribution of electronic states near the Fermi level, which modifies both the electron–phonon coupling and the effective Coulomb repulsion. When combined with proximity-induced pairing across the interface, these effects lead to an overall strengthening of superconductivity.

A key result of this work is the non-monotonic dependence of the critical temperature on layer thickness. This behavior reflects the presence of multiple characteristic confinement length scales associated with the different materials in the heterostructure. As the thickness is varied, the competition between confinement-enhanced pairing and proximity-induced coupling produces maxima in $T_c$, as well as regimes where superconductivity is either enhanced or suppressed. This provides a clear physical mechanism for tuning superconductivity through nanoscale structural design.

Another important outcome is the prediction of superconductivity in systems composed of materials that are not superconducting in bulk form. In particular, we showed that normal-metal/normal-metal bilayers can become superconducting when one of the layers develops confinement-induced superconductivity, which is then transferred to the second layer through proximity coupling. This mechanism opens a new route to designing superconductors from otherwise non-superconducting materials.

From a broader perspective, the present results demonstrate that superconductivity can be engineered not only by chemical composition, but also by geometric confinement and heterostructuring. This approach is particularly appealing because it does not rely on complex material synthesis or exotic pairing mechanisms, but instead exploits well-understood physical effects within conventional phonon-mediated superconductivity.

The predicted bilayer structures are experimentally accessible using standard thin-film growth techniques such as sputtering, molecular beam epitaxy, or atomic-layer deposition. Furthermore, the use of simple metals with nearly spherical Fermi surfaces ensures that the theoretical predictions are robust and based on well-established material parameters.

The present work also opens several promising directions for future research. These include the extension to multilayer superlattices, the investigation of asymmetric layer thicknesses, and the exploration of confinement effects in systems with stronger electron–phonon coupling. Another interesting direction is the study of transport properties, including critical currents and magnetic-field response, in confined heterostructures.

From an applications perspective, the ability to enhance and tune superconductivity through nanoscale design may be relevant for superconducting electronics, quantum computing, and nanoscale sensors. In particular, engineered heterostructures with enhanced critical temperatures could improve device performance and enable superconducting technologies based on simple metallic materials.

There are no particular difficulties in extending this theory to multiband material systems such as pnictides \cite{UmmaMulti1,UmmaMulti2,UmmaMulti3,UmmaMulti4}, except that these materials have Fermi surfaces that cannot be approximated by a sphere and therefore the quantum confinement theory must be generalized to more complicated shapes.

Overall, our results establish quantum confinement combined with proximity effects as a powerful and versatile tool for engineering superconductivity in nanoscale heterostructures.

\section*{Data availability} All data that support the findings of this study are included within the article.

%

\newpage

\bibliographystyle{apsrev4-2}
\bibliography{refs}

\end{document}